\documentclass[journal = nalefd]{achemso}
\usepackage{amsmath}
\usepackage{amssymb}
\usepackage{dcolumn} 
\usepackage{setspace}
\usepackage{array}
\usepackage{graphicx}
\usepackage{booktabs}
\usepackage{gensymb}
\usepackage{multirow}
\usepackage{psfrag}
\nocite{*} 

\newcommand{\eqn}[1]{\mbox{Eq.\hspace{1pt}(\ref{#1})}}
\newcommand{\eqs}[2]{\mbox{Eq.\hspace{1pt}(\ref{#1}--\ref{#2})}}

\newcommand{\eqtn}[2]{\begin{equation} \label{#1} #2 \end{equation}}

\def\brp{{\mathbf{r}^{\prime}}}

\def\deltarho{{\langle\Delta\rho\rangle}}

\def\br{{\mathbf{r}}}

\def\d{{\mathrm{d}}}
\def\rhor{{\rho({\bf r})}}

\def\rhoi{{\rho_I}}

\def\rhoir{{\rho_I({\bf r})}}

\def\rhojrp{{\rho_J({\bf r}^{\prime})}}
\def\sumi{{\sum_I^{N_S}}}

\author{Wenhui Mi}
\affiliation{Department of Chemistry, Rutgers University, Newark, NJ 07102, USA}
\alsoaffiliation{Department of Physics, Rutgers University, Newark, NJ 07102, USA}
\author{Michele Pavanello}
\email{m.pavanello@rutgers.edu}
\affiliation{Department of Chemistry, Rutgers University, Newark, NJ 07102, USA}
\alsoaffiliation{Department of Physics, Rutgers University, Newark, NJ 07102, USA}

\title{Nonlocal Subsystem Density Functional Theory}
\graphicspath{{./figures/}}

\begin{document}
\maketitle
\begin{abstract}
By invoking a divide-and-conquer strategy, subsystem DFT dramatically reduces the computational cost of large-scale, \textit{ab-initio} electronic structure simulations of molecules and materials. The central ingredient setting subsystem DFT apart from Kohn-Sham DFT is the non-additive kinetic energy functional (NAKE). Currently employed NAKEs are at most semilocal (i.e., they only depend on the electron density and its gradient), and as a result of this approximation, so far only systems composed of weakly interacting subsystems have been successfully tackled. In this work, we advance the state-of-the-art by introducing fully nonlocal NAKEs in subsystem DFT simulations for the first time. A benchmark analysis based on the S22-5 test set shows that nonlocal NAKEs considerably improve the computed interaction energies and electron density compared to commonly employed GGA NAKEs, especially when the inter-subsystem electron density overlap is high. Most importantly, we resolve the long standing problem of too attractive interaction energy curves typically resulting from the use of GGA NAKEs. 
\end{abstract}

\newpage
\textit{Ab-initio} models of realistically sized materials has become an ultimate goal for quantum chemistry and material science. To achieve this aim, recent years have witnessed the development of a variety of methods, such as density functional theory (DFT)\cite{kohn1965}, as well as multilevel/multiscale computational protocols such as QM/MM\cite{senn2009,shur2003}. Quantum embedding methods have recently gained fame and branched into several directions. Among them, subsystem DFT (sDFT) is becoming popular \cite{weso2015,gome2012,jaco2014,krish2015a,nafz2014}. The idea behind sDFT is the one of dividing the system into a set of interacting subsystems whose interaction is accounted for approximately in a way that leverages pure density functionals \cite{yang1990,huang2011,grit2013,good2010}. The simplicity of the algorithms involved and the propensity for massive parallelization has driven a number of implementations of sDFT methods in various mainstream quantum simulations codes \cite{jaco2008b,fderelease,Andermatt_2016}, and successfully applied to a vast array of chemical problems, for instance, structure and dynamics of molecular liquids \cite{Mi_2019,Genova_2016a}, solvent effects on different types of spectroscopy\cite{neug2005b,neug2010a}, magnetic properties \cite{jaco2006b,bulo2008,neug2005f,kevo2013,weso1999}, excited states\cite{neug2010,casi2004,pava2013b,neug2005b,garc2006,ramo2015c,Umerbekova_2018}, charge transfer states \cite{pava2013a,solo2014,ramos2015}, and bulk impurity models \cite{Toelle_2019}.

In sDFT, the total electron density, $\rhor$, is expressed as a sum of subsystem contributions. Namely,
\eqtn{sum}{\rhor = \sumi\rhoir,}
where $N_{s}$ is the total number of subsystems considered.
The electron density of each subsystem is obtained by variationally minimizing the total energy functional
\begin{align}
\nonumber
E[\{\rhoi\}] = & \sumi E[\rhoi,v_{ext}^I] + \\
\nonumber
            + &\underbrace{T_s[\rho]-\sumi T_s[\rhoi]}_{T_s^\text{nad}[\{\rhoi\}]} + \underbrace{E_{xc}[\rho]-\sumi E_{xc}[\rhoi]}_{E_{xc}^\text{nad}[\{\rhoi\}]} + \\
            + & \frac{1}{2}\sum_{I\neq J}^{N_s} \int \frac{\rhoir \rhojrp}{|\br-\brp|}\d\br\d\brp + \sum_{I\neq J}^{N_s} \int \rhoir v_{ext}^J(\br)\d\br,
\label{ef}
\end{align}
where $v_{ext}^J$ is the external potential associated with subsystem $J$, and by $\{\rhoi\}$ it is intended to indicate the collection of all subsystem densities. The subsystem energy functionals, $E[\rhoi,v_{ext}^I]$, are functionals of both, the subsystem external potentials and of the subsystem electron densities. The external potential is subsystem-additive (i.e., $v_{ext}(\br)=\sumi v_{ext}^I(\br)$).

Carrying out sDFT simulations involves solving one Kohn--Sham (KS) like equation for each subsystem whose KS potential, $v_{KS}(\br)$, is augmented by an embedding potential that accounts for the interactions with all other subsystems. Namely,
\begin{equation}
 \label{ksequ}
 \left[\frac{-\nabla^2}{2}+\upsilon^{I}_{KS}(\br)+\upsilon^{I}_{emb}(\br)\right]\phi_{i}^I(\br)=\epsilon_{i}^I(\br)\phi_{i}^I(\br),
\end{equation}
where $\phi_{i}^I(\br)$ and $\upsilon^{I}_{emb}(\br)$ are the KS wavefunctions and the embedding potential of subsystem $I$, respectively. The embedding potential can be written as follows\cite{jaco2014,krish2015a}:
\begin{align}
 \label{embpot}
 \nonumber
 \upsilon^{I}_{emb}(\br)=&\sum^{N_s}_{J\neq I}\left[\int \frac{\rho_J(\brp)}{|\br-\brp|}d\brp+\sum_{J} v_{ext}^J(\br)\right]+ \\
&+\frac{\delta T_{\rm s}^\text{nad}[\{\rhoi\}]}{\delta\rhoir}+\frac{\delta E^\text{nad}_{\rm xc}[\{\rhoi\}]}{\delta\rhoir}.
\end{align}
In the above, $T_{\rm s}$ and $E_{\rm xc}$ are kinetic energy density functionals (KEDF) and exchange--correlation (xc), respectively. 

In KS-DFT, $T_{s}[\rho]$ is evaluated exactly from the KS orbitals of the system. Conversely, in a sDFT scheme, approximate nonadditive kinetic energy functionals (NAKE, defined in \eqn{ef}) are employed. Employing NAKE constitutes the most important and crucial difference between carrying out a KS-DFT simulation and a sDFT simulation \cite{gotz2009,weso1996}. 

NAKEs are typically derived from semilocal KEDFs\cite{gotz2009} and have been at most of Laplacian level \cite{Laricchia_2013b}. However, it is common knowledge that semilocal NAKEs cannot approach a regime in which the subsystem electron densities strongly overlap where they typically give wrong interaction energy curves \cite{sinh2015,schl2015}. These limitations originate from the natural nonlocality of the underlying KEDF \cite{chac1985} and in turn of the NAKEs. In this work, we tackle these issues by employing state-of-the-art nonlocal KEDFs to generate NAKEs. 


Even though nonlocal KEDFs have a long history in OF-DFT simulations \cite{wesolowski2013recent,karasiev2012issues,witt2018orbital}, to the best of our knowledge they have not yet been employed as NAKEs.  This is probably because in sDFT, the distribution of  electron densities are usually more localized compared to the electron density of the supersystem \cite{Nafziger_2015,pava2011b,ramo2015b}. Thus, when developing nonlocal NAKEs, KEDFs must be able to correctly simulate both homogeneous and non-homogeneous systems, and  be numerically stable. 

The ability to approach inhomogeneous systems is the most challenging property to satisfy because the  nonlocal KEDFs have been historically developed for extended metallic systems whose electron density is close to uniform. The typical ansatz chosen for nonlocal functionals is:
\begin{equation}
\label{NL}
T_s[\rho] =\underbrace{T_{TF}[\rho] + T_{vW}[\rho]}_{T_{TV}[\rho]} + T_{NL}[\rho]
\end{equation}
where, $T_{TF}[\rho]$ is Thomas-Fermi (TF) functional \cite{fermi1927,thom1927}, $T_{vW}[\rho]$ is the von Weiz\"sacker (vW) functional \cite{weiz1935}, $T_{NL}[\rho]$ is the nonlocal part. The corresponding KEDF potential can be written as:
\begin{equation}
v_{T_{s}}(\br)=\frac{\delta T_{TV}[\rho]}{\delta \rho(\br)} +\frac{\delta T_{NL}[\rho]}{\delta \rho(\br)}=v_{TV} (\br)+ v_{NL}(\br),
\end{equation}
where $v_{TV}(\br)$ is the Thomas-Fermi-vW potential which we will later discuss.
 The nonlocal part is defined by a double integration of the electron density evaluated at two different points in space and an effective interaction, the so called kernel, $\omega$:
\begin{equation}
T_{NL}[\rho]=\int \int \rho^{\alpha}(\br)\omega[\rho](\br,\br')\rho^{\beta}(\br')d\br d\br'
\end{equation}   
where $\alpha$ and $\beta$ are positive numbers. The kernel is related to the second functional derivative of the KEDF with respect to the electron density \cite{wang2000} and is typically approximated by a function of only $|\br-\brp|$. 

The available nonlocal KEDFs\cite{wang1998,wang1999,huan2010,wang1992,mi2018nonlocal,Pearson_1993,smar1994,perr1994}, can be categorized in functionals whose kernel only depends on the average electron density (i.e., $\rho_0$ which is well defined only for condensed-phase systems), and functionals whose kernel instead depends on the total electron density and not just its average\cite{wang1992}. Clearly, nonlocal KEDFs with a density-independent kernel cannot be directly employed as NAKEs because the presence of the restrictive parameter, $\rho_{0}$, would make the KEDF unable to approach inhomogeneous systems. Unfortunately, some KEDFs with density dependent kernels are either too expensive (i.e., HC \cite{huan2010}) or numerically unstable for arbitrary inhomogeneous systems (i.e., WGC \cite{wang1999}), thus we will not employ them here. 

We recently proposed a new series of nonlocal KEDFs featuring local density dependent kernels which we showed \cite{mi2019LMGP} can predict accurately the electron density, energy and forces for clusters of metallic and group III and V atoms. These functionals are based on and generalize existing functionals with density independent kernels (such as WT \cite{wang1992}, MGPA and MGPG functionals\cite{mi2018nonlocal}). The generalization allows them to approach inhomogeneous systems  because they feature fully density dependent kernels\cite{mi2019LMGP}. 

Let us summarize the employed kernels, starting with the WT kernel \cite{wang1992} expressed in reciprocal space ($q$ is the reciprocal space variable for $|\br-\brp|$) and $\eta(q)=\frac{q}{2k_F}$ with $k_F$ being the Fermi wavevector),
\begin{align}
\label{WT}
\omega_{WT}(q)=\frac{6}{5}\frac{\pi^{2}}{(3\pi^{2})^{1/3}}G_{NL}(\eta(q))
\end{align}
which then is modified to satisfy functional integration relations \cite{mi2018nonlocal} by the addition of one correction term. Namely,
\begin{align}
\label{MGP}
\omega_{x,y}(q)=\omega_{WT}(q)-\frac{\pi^{2}x}{(3\pi^{2})^{1/3}}\int_{0}^{1}dt ~t^{y}\frac{d G_{NL}(\eta(q,t))}{dt}.
\end{align}
%
where
\begin{equation}
G_{NL}(\eta)=\left( \frac{1}{2} + \frac{1-\eta^2}{4\eta} \ln \left| \frac{1+\eta}{1-\eta} \right| \right)^{-1} - 3\eta^{2} -1
\end{equation}
and, MGP is given by $(x,y)=\left(\frac{6}{5},\frac{6}{5}\right)$, MGPA by $(x,y)=\left(\frac{3}{5},\frac{6}{5}\right)$ and MGPG by $(x,y)=\left(\frac{6}{5},\frac{3}{5}\right)$. The only difference between MGP/A/G is the way a kernel is symmetrized. We refer the interested reader to the supplementary information of Ref.\ \citenum{mi2018nonlocal}.

In Ref.\ \citenum{mi2019LMGP} we developed a technique to generalize WT,  MGP/A/G functionals to approach localized, finite systems by invoking spline techniques to obtain kernels no longer dependent only on the average electron density but instead they are dependent (locally) on the full electron density function. In this way, we generate the LWT, LMGP/A/G functionals from the kernels mentioned in \eqs{WT}{MGP}.

At implementation time, we noticed that the terms $T_{TV}=T_{TF}+T_{vW}$ and $T_{NL}$ each can lead to numerical instability for different reasons. The issue for $T_{TV}[\rho]$ originates from its quadratic dependence on the density gradient. In the typical GGA formalism:
\begin{equation}
T_{TV}[\rho]=\int t_{TF}(\br) F_{TV}(s(\br)) d\br 
\end{equation}
where $s$ is the  dimensionless reduced density gradient,  $s=\frac{|\nabla \rho|}{2\rho k_{f}}=\frac{1}{2(3\pi^{2})^{1/3}}\frac{|\nabla \rho|}{\rho^{3/4}}$, the enhancement  factor $F_{TV}(s)=1+\frac{5}{3}s^{2}$, and $t_{TF}(\br) = \frac{3}{10}(3\pi^2)^{\frac{2}{3}}\rho^{\frac{5}{3}}(\br)$. Numerical inaccuracies arise at large $s$ because in this limit, $F_{TV}(s)$ is unbound and the error in the density becomes uncontrollable.

\begin{figure}
\includegraphics[width=0.8\textwidth]{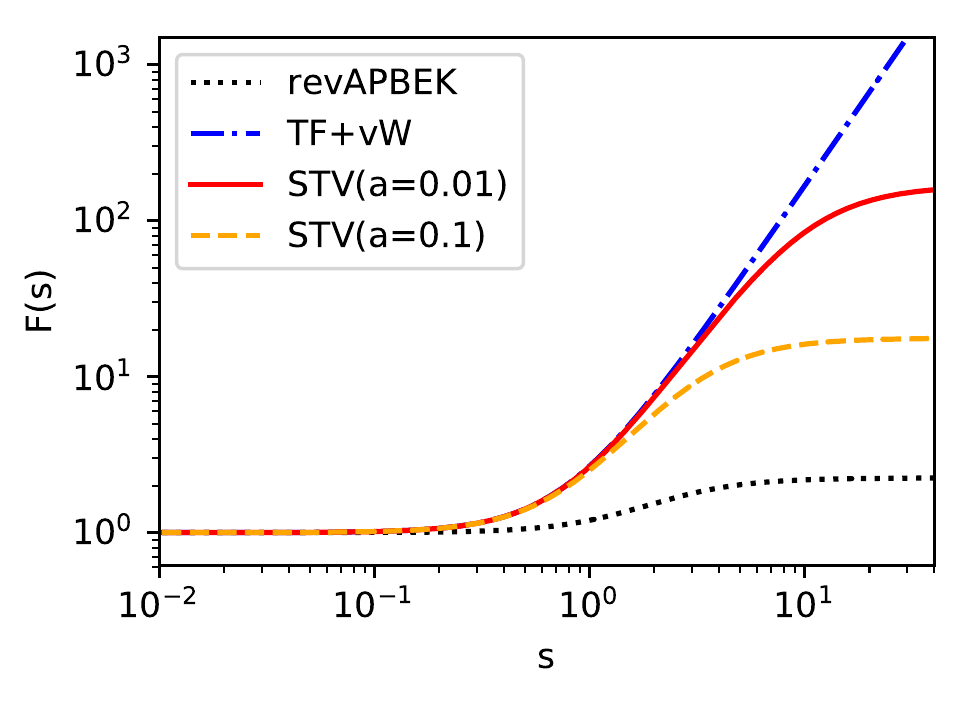}
\caption{\label{FS}Enhancement factors for smooth version of the $T_{TV}$ functional (STV) (a = 0.01 red solid), STV(a =0.1 orange dashline), pure $T_{TV}$ (blue dotdash), and revAPBEK (black dots).}
\end{figure}

Thus, we need to find a proper way to cap $F_{TV}(s)$ for large $s$. To achieve this aim, we borrow a formalism similar to PBE exchange \cite{PBEc} and reshape the enhancement factor of  Thomas-Fermi (TF) plus von Weizs\"acker (vW) kinetic energy functional in a stable formalism (named STV): 
\begin{equation}
F_{STV}(s) = 1.0 + \frac{5}{3}\frac{s^{2}}{1.0+as^{2}}. 
\end{equation}
In this formalism, when $a$=0, $F_{STV}(s)$ is same as the original $F_{TV}$;  by increasing $a$, $F_{STV}$ smoothly approaches to a constant number for large $s$, which should ameliorate the numerical inaccuracies. Fig.\ref{FS}  compares STV functionals (for both a=0.1 and 0.01) with the $T_{TV}$ and revAPBEK enhancement factors. 

In addition to the numerical problem for the $T_{TV}$ KEDF, the nonlocal KEDF potentials also need to be carefully implemented in the low electron density regions. The nonlocal kinetic potentials for all the nonlocal functionals considered in this work share the form:  
\begin{equation}
\label{nl2}
v_{NL}(\br)=\rho^{-1/6}(\br)F^{-1}\bigg[\tilde\rho^{5/6}(\mathbf{q})\omega(q)\bigg](\br),
\end{equation}
where $\tilde\rho^{5/6}(\mathbf{q})=F\big[\rho^{5/6}(\br)\big](\mathbf{q})$, $\omega(q)$ is the nonlocal kernel expressed in reciprocal space, $F$ and $F^{-1}$ represent the fast Fourier transform and inverse fast Fourier transform, respectively. In \eqn{nl2} it is made clear that we approximate the real-space kernel as a function of only $|\br-\brp|$ resulting in a dependence on only the magnitude of the reciprocal space vector $q=|\mathbf{q}|$. In the same equation there is a $\rho^{-1/6}(\br)$ prefactor, which is numerically noisy in the low electron density regions. To eliminate this issue, a local density weighted mix of GGA and nonlocal kinetic potential scheme is proposed: 
\begin{equation}
\label{gga}
v_{T_s}[\rho](\br) = \bigg(v_{NL}[\rho](\br) + v_{STV}[\rho](\br)\bigg) W[\rho](\br) + v_{GGA}[\rho](\br) \bigg(1-W[\rho](\br)\bigg)
\end{equation}  
 where $W[\rho](\br)=\frac{\rho(\br)}{\rho_{max}}$, $\rho_{max}$ is the maximum value of electron density in the system, and $v_{GGA}$ is the KEDF potential from a GGA functional (here we choose revAPBEK). In this way, for the region of space with low electron density, the kinetic potential is mainly contributed by the GGA functional instead of the nonlocal part. The procedure in \eqn{gga} cures the numerical instability of the nonlocal part of the potential.

With the KEDF potential in hand, the kinetic energy can be evaluated by line integration:
\begin{equation}
T_{s}[\rho]=\int \rho(\br) d\br \int_{0}^{1} v_{T_s}[\rho_{t}](\br) dt
\end{equation}
where $\rho_{t}(\br)=t \rho(\br)$.


We now present pilot calculations aimed at assessing  the performance of our newly proposed nonlocal NAKEs based on the following KEDFs: LWT, LMGPA, LMGPG. We select the S22-5 test set (non-covalently interacting complexes at equilibrium and displaced geometries \cite{Grafova_2010}) as benchmarks. The molecules are placed in an orthorhombic (cubic) box where the periodic boundary condition is applied. The separations between the studied molecules and their nearest-neighbor periodic images are at least 12\AA. This is a large enough separation to ensure that spurious self-interactions are negligible. Both our new proposed nonlocal NAKEs and the GGA functionals have been implemented in a development version of the embedded Quantum ESPRESSO (eQE) package\cite{fderelease}. All KS-DFT benchmark calculations are performed with the Quantum ESPRESSO (QE) package\cite{qe}. In both subsystem DFT and KS-DFT calculations, the Perdew-Burke-Ernzerhof (PBE) form of the GGA xc functional\cite{PBEc} is employed. In order to show the influence of the xc functional on the results, the nonlocal rVV10\cite{rvv10} functional is also adopted. Ultrasoft  pseudopotentials are adopted\cite{vand1990} (specifically the GBRV version 1.4 \cite{garr2014}). The plane wave cutoffs are 70 Ry and 400 Ry, for the wave functions and density, respectively.

When comparing the interaction energies summarized in Figure S1 of the supplementary materials \cite{epaps}, both revAPBEK and LMGPA functional reproduce the benchmark within 2 kcal/mol for weakly interacting systems. Decreasing the separation between two fragments from S22(2.0) to S22(0.9), (S22($x$) indicates that the distance between the two fragments is given by $x \times R_{eq}$, where $R_{eq}$ is the equilibrium distance as computed by couple cluster theory) as by expectations increases the deviation of sDFT and KS-DFT interaction energies. To clearly show the performance of LMGPA and the revAPBEK functionals for strongly interacting configurations, here we focus on the interaction energies and total electron densities (i.e., the sum of the two subsystems' densities for sDFT and the total density for KS-DFT) computed for the S22(0.9) case. Results for all other systems are provided in the supplementary information document \cite{epaps}. 
\begin{figure}
\includegraphics[width=0.9\textwidth]{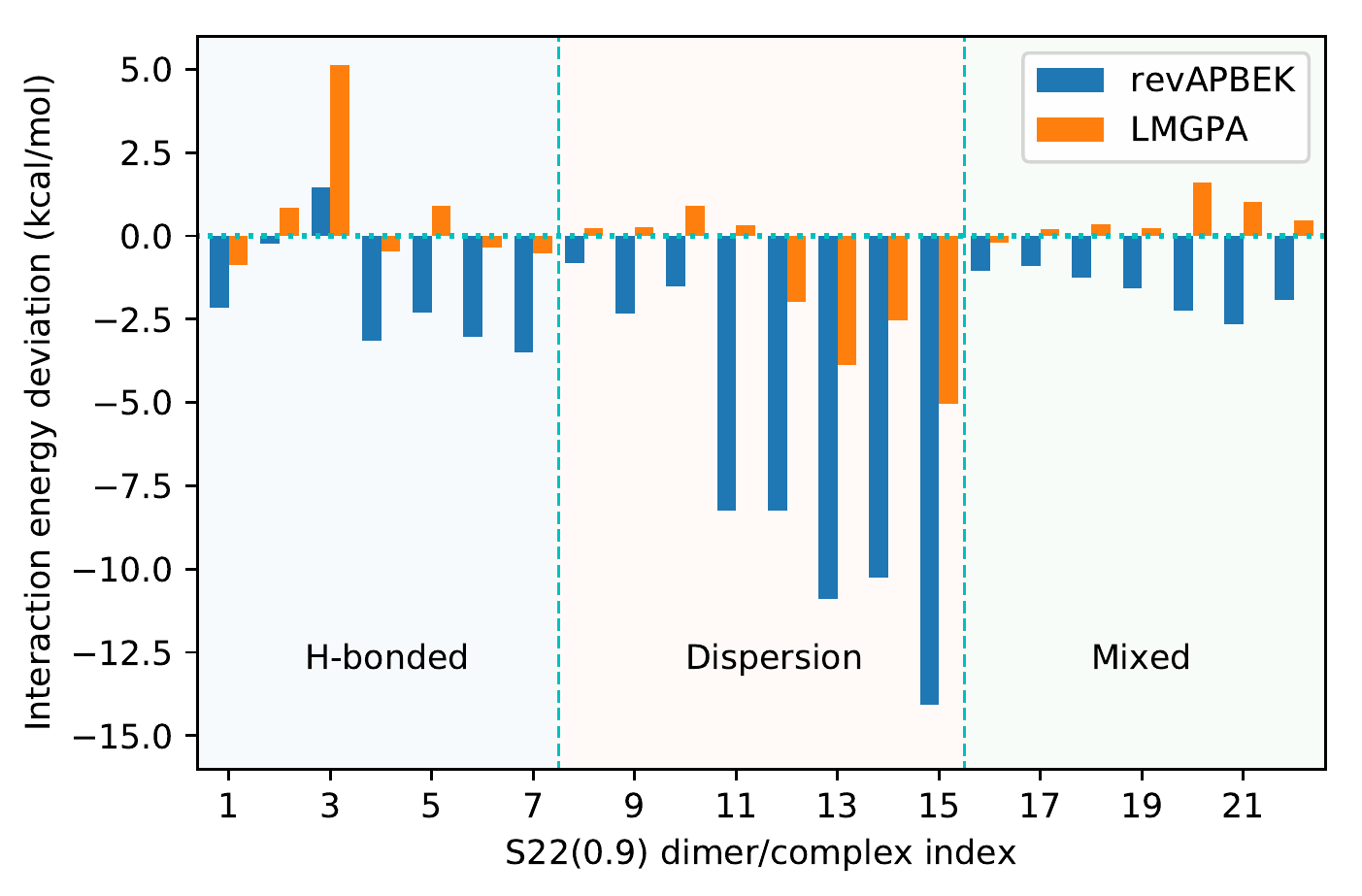}
\caption{\label{Energy_22} Interaction energy deviations in kcal/mol obtained by sDFT with revAPBEK and LMGPA NAKE functionals compared with corresponding KS-DFT results. All geometries from the S22(0.9) set and  the corresponding complexes of these indexes are listed in Table S1 of support information.}
\end{figure}
Figure \ref{Energy_22} shows that the LMGPA functional considerably improves the revAPBEK results for all systems with a max deviation of the interaction energy of about 5 kcal/mol. This compares quite well against more than 14 kcal/mol for revAPBEK. The only exception is formic acid dimer in which the two fragments are bonded by a double hydrogen bond. The abnormality of this dimer is revealed in two aspects:  it is the only case where the revAPBEK functional overestimates the total energy, and it also the only case where the revAPBEK functional performs better than LMGPA. This system is also associated with the largest electron density deviation (\textit{vide infra}) and thus the revAPBEK apparent good performance is due to fortuitous error cancellation.  

To further quantify the performance of the LMGPA NAKE functional, we summarize the root-mean-square deviations (RMSD) of the interaction energies sDFT with different NAKEs (revAPBEK and LMGPA) from the reference KS-DFT results are showed in Table \ref{Energy}. Inspecting the table, it is clear that LMGPA outperforms revAPBEK. The total RMSD for LMGPA is just 1.97 kcal/mol which is within the chemical accuracy, while revAPBEK results in a RSMD of 5.36 kcal/mol or about three times larger than the LMGPA RMSD. We notice that the LMGPA particularly improves the dispersion bound systems for which it obtains much improved results (2.54 kcal/mol) compared to revAPBEK (8.42kcal/mol). Moreover, the long standing issue of GGA NAKEs that generate too attractive interaction energy curves (which is also clear from Figure \ref{Energy_22}) is cured by the LMGPA NAKE functional. 

\begin{table}[htp]
\begin{center}
\begin{tabular}{lcccc}
 \toprule

NAKEs            & Hydrogen & Dispersion & Mixed & Total \\
 \hline
revAPBEK              &2.49            & 8.42             & 1.76     & 5.36  \\
LMGPA                         &2.05            & 2.54             & 0.77     & 1.97   \\
 \bottomrule
\end{tabular}
\end{center}
\caption{Summary of the root-mean-square deviations (RMSD) of the interaction energies computed with sDFT carried out with revAPBEK and LMGPA NAKEs and the PBE xc functional compared to the reference KS-DFT results. All geometries are from the S22(0.9) set and RMSD are in kcal/mol.}
\label{Energy}
\end{table}

\begin{figure}
\includegraphics[width=1.0\textwidth]{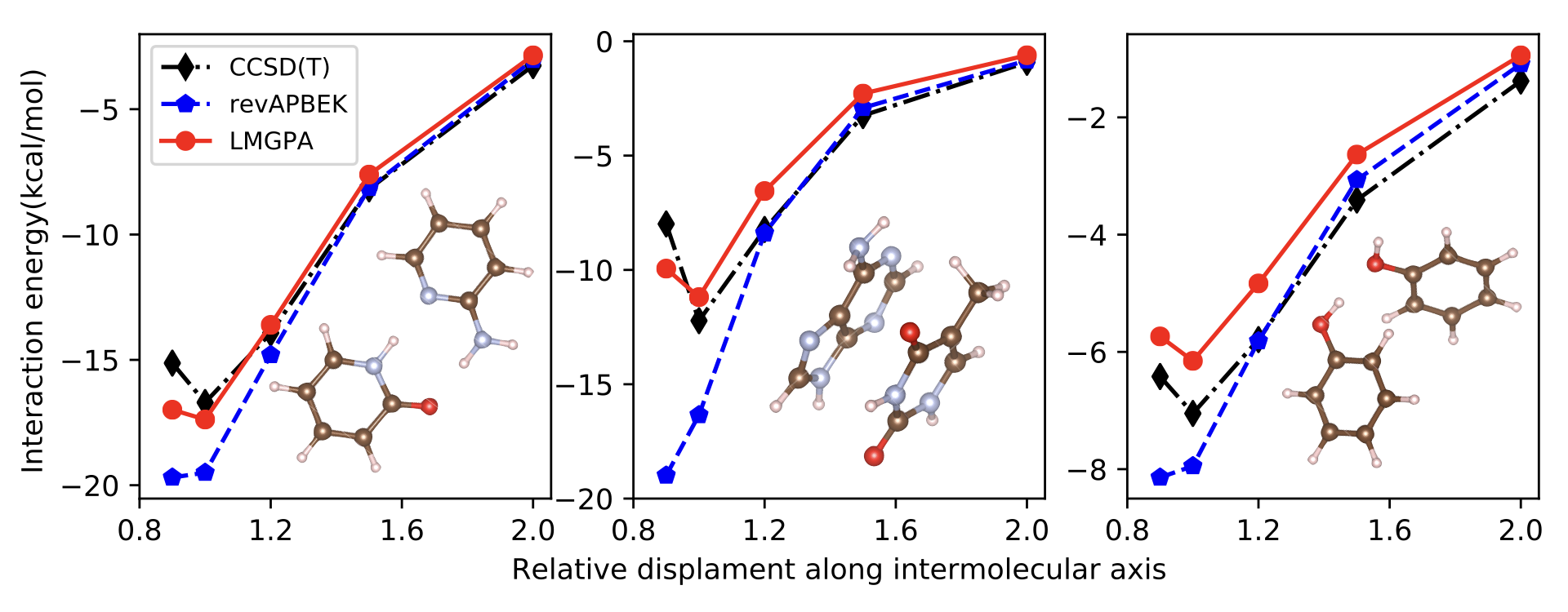}
\caption{\label{EOS} Interaction energies obtained by sDFT with revAPBEK and LMGPA NAKEs in conjunction with rVV10 xc functional compared against CCSD(T).}
\end{figure}

It is now clear that LMGPA delivers good interaction energies with sub-chemical accuracy deviations from KS-DFT. We wish to test its ability to deliver accurate interaction energies in comparison to the benchmark CCSD(T) energies\cite{Grafova_2010,CCSDT}. In a previous formal work by our group \cite{sinh2015} we showed that once sDFT is associated with an exact $T_s^\text{nad}$ and a nonlocal xc functional, interaction energies become closer to benchmark results. Thus, here we compare LMGPA and revAPBEK NAKEs in conjunction with the rVV10 xc functional.  Due to its nonlocal nature, rVV10 has been shown to be much more reliable than GGA xc functionals in KS-DFT calculations \cite{vydr2011}, especially for the dispersion bonded systems. In this work, we witness a similar outcome as evident from  the benchmarks for each type of bonding systems showed in Figure \ref{EOS}.  KS-DFT with both PBE and rVV10 xc functionals are available in the supporting information section \cite{epaps}. As showed in Figure S4-5, KS-DFT with rVV10 functional can obtain nearly exactly the same results as CCSD(T) for all systems. Figure \ref{EOS} indicates that in line with the results presented above, LMGPA obtains correct equilibrium bonding length and the order of energies. This is a major improvement in comparison to the revAPBEK results which feature a well characterized deficiency of too attractive energy curves \cite{sinh2015,schl2015,Mi_2019}. Moreover, in order to show the influence of the choice of xc functionals on the sDFT performance, we benchmarked the sDFT interaction energy deviations from the corresponding KS-DFT results. As showed in Figure S6, the sDFT results are nearly independent from the choice of xc functional, further reinforcing the conclusion that the nonlocal LMGPA functional resolved the long standing problem of too attractive energy curves computed by semilocal NAKEs.     

Reproducing the electron density is also important in evaluation of the performance of functionals \cite{Medvedev_2017,Wasserman_2017}. Thus, a more insightful comparison is made by calculating the number of electrons misplaced by sDFT, $\deltarho$, defined as:
\begin{equation}
\label{drho}
\deltarho =\frac{1}{2}\int |\rho_{sDFT}(\br)-\rho_{KS}(\br)| d\br
\end{equation}  
This value is an important quantity, as it vanishes only when sDFT and KS-DFT electron densities coincide. The RMSD of $\deltarho$ for revAPBEK and LMGPA NAKEs results are showed in Table \ref{Density_tot}.  

\begin{table}[htp]
\begin{center}
\begin{tabular}{lccccc}
 \toprule
     r/r$_0$  & 0.9 & 1.0 & 1.2 & 1.5 &2.0  \\
 \hline      
revAPBEK &0.0600 & 0.0370 & 0.0148 & 0.0042 & 0.0008\\
LMGPA &0.0583 & 0.0361 & 0.0148 & 0.0042 & 0.0008\\   
 \bottomrule
\end{tabular}
\end{center}
\caption{RMSD of the 22 $\deltarho$ obtained by sDFT with different NAKEs (revAPBEK and LMGPA) for equilibrium and additional four nonequilibrium geometries of the S22-5 test set. }
\label{Density_tot}
\end{table}
As expected, when the interaction between the two subsystems transitions from weakly to strong (corresponding to from S22(2.0) to S22(0.9)), the $\deltarho$ value is also increases. For the S22(0.9) and S22(1.0) sets, LMGPA performs slightly better than revAPBEK.  Since in the weakly interaction regime for both NAKEs can generate nearly the same and accurate electron density, we will just focus on the set with the strongest interactions (i.e., the S22(0.9)). 

\begin{figure}
\includegraphics[width=1.0\textwidth]{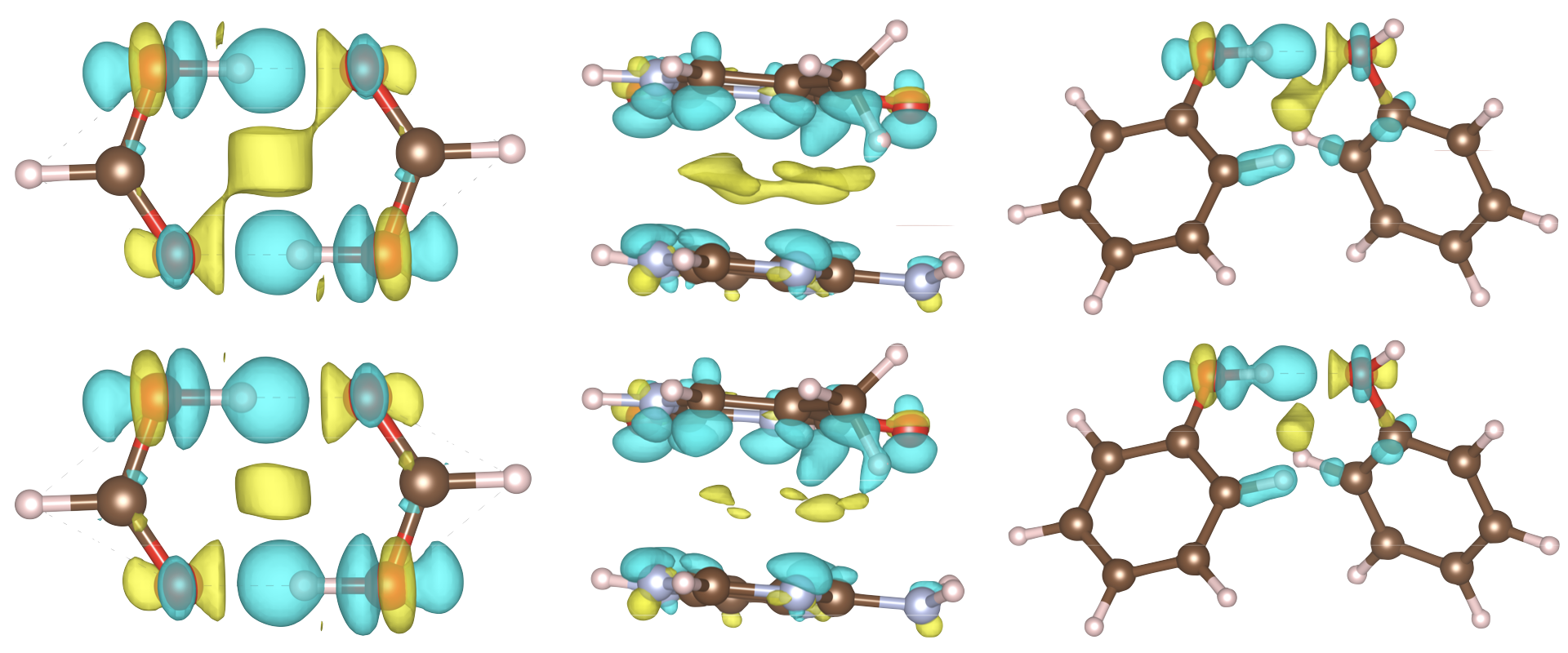}
\caption{\label{DD} $\langle\rho\rangle$ obtained by sDFT with revAPBEK (top) and LMGPA (below) NAKEs compared with corresponding KS-DFT results. Isosurfaces of, L-to-R: $1.0e-3$, $5.0e-4$, $5.0e-4$.}
\end{figure}

\begin{table}[htp]
\begin{center}
\begin{tabular}{llcccc}
 \toprule

Bond type  & Hydrogen & Dispersion & Mixed & Total  \\
 \hline                
revAPBEK &0.0805 & 0.0600 & 0.0270 & 0.0600\\
LMGPA      &0.0801 & 0.0561 & 0.0261 & 0.0583\\
 \bottomrule
\end{tabular}
\end{center}
\caption{RMSD for $\deltarho$ defined in \eqn{drho} for different bonding types in the S22 (0.9) set. }
\label{Density}
\end{table}
The results for each bonding type is summarized in Table \ref{Density}. Compared with revAPBEK, LMGPA NAKE obtains smaller $\deltarho$ for all cases indicating that it can generate more accurate electron density for all types of bonding.  We select the complexes which generate the largest $\deltarho$ for each type of bonding and plot the corresponding isosurface plots of density difference (compared with KS-DFT) for sDFT with revAPBEK and LMGPA, see Figure \ref{DD}.  As expected, the density difference mainly occurs on the overlap regions between the two subsystems. We now have a visual of the fact that LMGPA can generate more accurate electron densities compared to revAPBEK, since the density difference region is  much smaller than the revAPBEK results.

In the previous analysis, we just focus on LMGPA with $a=0.01$ in the definition of the smooth Thomas-Fermi-von Weizs\"acker, STV, functional. To benchmark the influence of the choice of $a$ and the performance of each kernel, both $a=0.01$ together with $a=0.1$ and other functionals (LWT and LMGPG) are also compared in the supporting information section \cite{epaps}. As shown in Figure S2, all these nonlocal functionals result in improved interaction energies compared against revAPBEK. In terms of electron density, Figures S7 and S8 show that all of the new nonlocal NAKEs obtain better results than revAPBEK. 
     
In conclusion, for the first time we employed nonlocal nonadditive Kinetic Energy functionals in subsystem DFT simulations. Our approach relies on  (1) adopting latest-generation nonlocal functionals featuring a fully density dependent kernel, correctly tackling systems with localized and inhomogeneous electron density;  (2) suppressing numerical instabilities in the evaluation of the von Weizs\"acker KEDF and nonlocal KEDF in the low electron density regions. Our approach leads to numerically stable and accurate subsystem DFT simulations. Benchmark tests against the well-known S22-5 test set indicate that our new approach not only can reproduce accurate interaction energies across bonding types (hydrogen, dispersion and mixed), but we also better reproduce the benchmark electron density. In addition, the new nonlocal subsystem DFT approach (that includes nonlocal NAKE and nonlocal xc functional) obtains correct equilibrium bonding lengths and correct shape of the energy curves compared to CCSD(T) energy curves, which have been a long standing challenge for semilocal sDFT.

\section{Acknowledgements}
We gratefully acknowledge discussions with Dr. Pablo Ramos and Dr. Xuecheng Shao. This material is based upon work supported by the National Science Foundation under Grant No. CHE-1553993. The authors acknowledge the Office of Advanced Research Computing (OARC) at Rutgers, The State University of New Jersey for providing access to the Amarel cluster and associated research computing resources that have contributed to the results reported here. URL: http://oarc.rutgers.edu 

\bibliography{./prg.bib}

\providecommand{\latin}[1]{#1}
\providecommand*\mcitethebibliography{\thebibliography}
\csname @ifundefined\endcsname{endmcitethebibliography}
  {\let\endmcitethebibliography\endthebibliography}{}
\begin{mcitethebibliography}{70}
\providecommand*\natexlab[1]{#1}
\providecommand*\mciteSetBstSublistMode[1]{}
\providecommand*\mciteSetBstMaxWidthForm[2]{}
\providecommand*\mciteBstWouldAddEndPuncttrue
  {\def\EndOfBibitem{\unskip.}}
\providecommand*\mciteBstWouldAddEndPunctfalse
  {\let\EndOfBibitem\relax}
\providecommand*\mciteSetBstMidEndSepPunct[3]{}
\providecommand*\mciteSetBstSublistLabelBeginEnd[3]{}
\providecommand*\EndOfBibitem{}
\mciteSetBstSublistMode{f}
\mciteSetBstMaxWidthForm{subitem}{(\alph{mcitesubitemcount})}
\mciteSetBstSublistLabelBeginEnd
  {\mcitemaxwidthsubitemform\space}
  {\relax}
  {\relax}

\bibitem[{Kohn} and {Sham}(1965){Kohn}, and {Sham}]{kohn1965}
{Kohn},~W.; {Sham},~L.~J. \emph{Phys. Rev.} \textbf{1965}, \emph{140},
  1133--1138\relax
\mciteBstWouldAddEndPuncttrue
\mciteSetBstMidEndSepPunct{\mcitedefaultmidpunct}
{\mcitedefaultendpunct}{\mcitedefaultseppunct}\relax
\EndOfBibitem
\bibitem[Senn and Thiel(2009)Senn, and Thiel]{senn2009}
Senn,~H.~M.; Thiel,~W. \emph{Angew. Chem. Int. Ed.} \textbf{2009}, \emph{48},
  1198--1229\relax
\mciteBstWouldAddEndPuncttrue
\mciteSetBstMidEndSepPunct{\mcitedefaultmidpunct}
{\mcitedefaultendpunct}{\mcitedefaultseppunct}\relax
\EndOfBibitem
\bibitem[Shurki and Warshel(2003)Shurki, and Warshel]{shur2003}
Shurki,~A.; Warshel,~A. \emph{Adv. Protein Chem.} \textbf{2003}, \emph{66},
  249--313\relax
\mciteBstWouldAddEndPuncttrue
\mciteSetBstMidEndSepPunct{\mcitedefaultmidpunct}
{\mcitedefaultendpunct}{\mcitedefaultseppunct}\relax
\EndOfBibitem
\bibitem[Wesolowski \latin{et~al.}(2015)Wesolowski, Shedge, and Zhou]{weso2015}
Wesolowski,~T.~A.; Shedge,~S.; Zhou,~X. \emph{{Chem. Rev.}} \textbf{2015},
  \emph{115}, 5891--5928\relax
\mciteBstWouldAddEndPuncttrue
\mciteSetBstMidEndSepPunct{\mcitedefaultmidpunct}
{\mcitedefaultendpunct}{\mcitedefaultseppunct}\relax
\EndOfBibitem
\bibitem[Gomes and Jacob(2012)Gomes, and Jacob]{gome2012}
Gomes,~A. S.~P.; Jacob,~C.~R. \emph{Annu. Rep. Prog. Chem., Sect. C: Phys.
  Chem.} \textbf{2012}, \emph{108}, 222--277\relax
\mciteBstWouldAddEndPuncttrue
\mciteSetBstMidEndSepPunct{\mcitedefaultmidpunct}
{\mcitedefaultendpunct}{\mcitedefaultseppunct}\relax
\EndOfBibitem
\bibitem[Jacob and Neugebauer(2014)Jacob, and Neugebauer]{jaco2014}
Jacob,~C.~R.; Neugebauer,~J. \emph{WIREs: Comput. Mol. Sci.} \textbf{2014},
  \emph{4}, 325--362\relax
\mciteBstWouldAddEndPuncttrue
\mciteSetBstMidEndSepPunct{\mcitedefaultmidpunct}
{\mcitedefaultendpunct}{\mcitedefaultseppunct}\relax
\EndOfBibitem
\bibitem[Krishtal \latin{et~al.}(2015)Krishtal, Sinha, Genova, and
  Pavanello]{krish2015a}
Krishtal,~A.; Sinha,~D.; Genova,~A.; Pavanello,~M. \emph{{J. Phys.: Condens.
  Matter}} \textbf{2015}, \emph{27}, 183202\relax
\mciteBstWouldAddEndPuncttrue
\mciteSetBstMidEndSepPunct{\mcitedefaultmidpunct}
{\mcitedefaultendpunct}{\mcitedefaultseppunct}\relax
\EndOfBibitem
\bibitem[Nafziger and Wasserman(2014)Nafziger, and Wasserman]{nafz2014}
Nafziger,~J.; Wasserman,~A. \emph{{J. Phys. Chem. A}} \textbf{2014},
  \emph{118}, 7623--7639\relax
\mciteBstWouldAddEndPuncttrue
\mciteSetBstMidEndSepPunct{\mcitedefaultmidpunct}
{\mcitedefaultendpunct}{\mcitedefaultseppunct}\relax
\EndOfBibitem
\bibitem[Yang(1991)]{yang1990}
Yang,~W. \emph{{Phys. Rev. Lett.}} \textbf{1991}, \emph{66}, 1438--1441\relax
\mciteBstWouldAddEndPuncttrue
\mciteSetBstMidEndSepPunct{\mcitedefaultmidpunct}
{\mcitedefaultendpunct}{\mcitedefaultseppunct}\relax
\EndOfBibitem
\bibitem[Huang \latin{et~al.}(2011)Huang, Pavone, and Carter]{huang2011}
Huang,~C.; Pavone,~M.; Carter,~E.~A. \emph{{J. Chem. Phys.}} \textbf{2011},
  \emph{134}, 154110\relax
\mciteBstWouldAddEndPuncttrue
\mciteSetBstMidEndSepPunct{\mcitedefaultmidpunct}
{\mcitedefaultendpunct}{\mcitedefaultseppunct}\relax
\EndOfBibitem
\bibitem[Gritsenko(2013)]{grit2013}
Gritsenko,~O.~V. In \emph{{Recent Advances in Orbital-Free Density Functional
  Theory}}; Wesolowski,~T.~A., Wang,~Y.~A., Eds.; World Scientific: Singapore,
  2013; Chapter 12, pp 355--365\relax
\mciteBstWouldAddEndPuncttrue
\mciteSetBstMidEndSepPunct{\mcitedefaultmidpunct}
{\mcitedefaultendpunct}{\mcitedefaultseppunct}\relax
\EndOfBibitem
\bibitem[Goodpaster \latin{et~al.}(2010)Goodpaster, Ananth, Manby, and {Miller,
  III}]{good2010}
Goodpaster,~J.~D.; Ananth,~N.; Manby,~F.~R.; {Miller, III},~T.~F. \emph{{J.
  Chem. Phys.}} \textbf{2010}, \emph{133}, 084103\relax
\mciteBstWouldAddEndPuncttrue
\mciteSetBstMidEndSepPunct{\mcitedefaultmidpunct}
{\mcitedefaultendpunct}{\mcitedefaultseppunct}\relax
\EndOfBibitem
\bibitem[Jacob \latin{et~al.}(2008)Jacob, Neugebauer, and Visscher]{jaco2008b}
Jacob,~C.~R.; Neugebauer,~J.; Visscher,~L. \emph{{J. Comput. Chem.}}
  \textbf{2008}, \emph{29}, 1011--1018\relax
\mciteBstWouldAddEndPuncttrue
\mciteSetBstMidEndSepPunct{\mcitedefaultmidpunct}
{\mcitedefaultendpunct}{\mcitedefaultseppunct}\relax
\EndOfBibitem
\bibitem[Genova \latin{et~al.}(2017)Genova, Ceresoli, Krishtal, Andreussi,
  {DiStasio Jr.}, and Pavanello]{fderelease}
Genova,~A.; Ceresoli,~D.; Krishtal,~A.; Andreussi,~O.; {DiStasio Jr.},~R.;
  Pavanello,~M. \emph{{Int. J. Quantum Chem.}} \textbf{2017}, \emph{117},
  e25401\relax
\mciteBstWouldAddEndPuncttrue
\mciteSetBstMidEndSepPunct{\mcitedefaultmidpunct}
{\mcitedefaultendpunct}{\mcitedefaultseppunct}\relax
\EndOfBibitem
\bibitem[Andermatt \latin{et~al.}(2016)Andermatt, Cha, Schiffmann, and
  VandeVondele]{Andermatt_2016}
Andermatt,~S.; Cha,~J.; Schiffmann,~F.; VandeVondele,~J. \emph{J. Chem. Theory
  Comput.} \textbf{2016}, \emph{12}, 3214--3227\relax
\mciteBstWouldAddEndPuncttrue
\mciteSetBstMidEndSepPunct{\mcitedefaultmidpunct}
{\mcitedefaultendpunct}{\mcitedefaultseppunct}\relax
\EndOfBibitem
\bibitem[Mi \latin{et~al.}(2019)Mi, Ramos, Maranhao, and Pavanello]{Mi_2019}
Mi,~W.; Ramos,~P.; Maranhao,~J.; Pavanello,~M. \emph{{J. Phys. Chem. Lett.}}
  \textbf{2019}, Submitted. available at arxiv.org/abs/1910.07359\relax
\mciteBstWouldAddEndPuncttrue
\mciteSetBstMidEndSepPunct{\mcitedefaultmidpunct}
{\mcitedefaultendpunct}{\mcitedefaultseppunct}\relax
\EndOfBibitem
\bibitem[Genova \latin{et~al.}(2016)Genova, Ceresoli, and
  Pavanello]{Genova_2016a}
Genova,~A.; Ceresoli,~D.; Pavanello,~M. \emph{{J. Chem. Phys.}} \textbf{2016},
  \emph{144}, 234105\relax
\mciteBstWouldAddEndPuncttrue
\mciteSetBstMidEndSepPunct{\mcitedefaultmidpunct}
{\mcitedefaultendpunct}{\mcitedefaultseppunct}\relax
\EndOfBibitem
\bibitem[Neugebauer \latin{et~al.}(2005)Neugebauer, Louwerse, Baerends, and
  Wesolowski]{neug2005b}
Neugebauer,~J.; Louwerse,~M.~J.; Baerends,~E.~J.; Wesolowski,~T.~A. \emph{{J.
  Chem. Phys.}} \textbf{2005}, \emph{122}, 094115\relax
\mciteBstWouldAddEndPuncttrue
\mciteSetBstMidEndSepPunct{\mcitedefaultmidpunct}
{\mcitedefaultendpunct}{\mcitedefaultseppunct}\relax
\EndOfBibitem
\bibitem[Neugebauer(2010)]{neug2010a}
Neugebauer,~J. \emph{Phys. Rep.} \textbf{2010}, \emph{489}, 1--87\relax
\mciteBstWouldAddEndPuncttrue
\mciteSetBstMidEndSepPunct{\mcitedefaultmidpunct}
{\mcitedefaultendpunct}{\mcitedefaultseppunct}\relax
\EndOfBibitem
\bibitem[Jacob and Visscher(2006)Jacob, and Visscher]{jaco2006b}
Jacob,~C.~R.; Visscher,~L. \emph{{J. Chem. Phys.}} \textbf{2006}, \emph{125},
  194104\relax
\mciteBstWouldAddEndPuncttrue
\mciteSetBstMidEndSepPunct{\mcitedefaultmidpunct}
{\mcitedefaultendpunct}{\mcitedefaultseppunct}\relax
\EndOfBibitem
\bibitem[Bulo \latin{et~al.}(2008)Bulo, Jacob, and Visscher]{bulo2008}
Bulo,~R.~E.; Jacob,~C.~R.; Visscher,~L. \emph{{J. Phys. Chem. A}}
  \textbf{2008}, \emph{112}, 2640--2647\relax
\mciteBstWouldAddEndPuncttrue
\mciteSetBstMidEndSepPunct{\mcitedefaultmidpunct}
{\mcitedefaultendpunct}{\mcitedefaultseppunct}\relax
\EndOfBibitem
\bibitem[Neugebauer \latin{et~al.}(2005)Neugebauer, Louwerse, Belanzoni,
  Wesolowski, and Baerends]{neug2005f}
Neugebauer,~J.; Louwerse,~M.~J.; Belanzoni,~P.; Wesolowski,~T.~A.;
  Baerends,~E.~J. \emph{{J. Chem. Phys.}} \textbf{2005}, \emph{123},
  114101\relax
\mciteBstWouldAddEndPuncttrue
\mciteSetBstMidEndSepPunct{\mcitedefaultmidpunct}
{\mcitedefaultendpunct}{\mcitedefaultseppunct}\relax
\EndOfBibitem
\bibitem[Kevorkyants \latin{et~al.}(2013)Kevorkyants, Wang, Close, and
  Pavanello]{kevo2013}
Kevorkyants,~R.; Wang,~X.; Close,~D.~M.; Pavanello,~M. \emph{{J. Phys. Chem.
  B}} \textbf{2013}, \emph{117}, 13967--13974\relax
\mciteBstWouldAddEndPuncttrue
\mciteSetBstMidEndSepPunct{\mcitedefaultmidpunct}
{\mcitedefaultendpunct}{\mcitedefaultseppunct}\relax
\EndOfBibitem
\bibitem[Wesolowski(1999)]{weso1999}
Wesolowski,~T.~A. \emph{{Chem. Phys. Lett.}} \textbf{1999}, \emph{311},
  87--92\relax
\mciteBstWouldAddEndPuncttrue
\mciteSetBstMidEndSepPunct{\mcitedefaultmidpunct}
{\mcitedefaultendpunct}{\mcitedefaultseppunct}\relax
\EndOfBibitem
\bibitem[Neugebauer \latin{et~al.}(2010)Neugebauer, Curutchet, Munioz-Losa, and
  Mennucci]{neug2010}
Neugebauer,~J.; Curutchet,~C.; Munioz-Losa,~A.; Mennucci,~B. \emph{{J. Chem.
  Theory Comput.}} \textbf{2010}, \emph{6}, 1843--1851\relax
\mciteBstWouldAddEndPuncttrue
\mciteSetBstMidEndSepPunct{\mcitedefaultmidpunct}
{\mcitedefaultendpunct}{\mcitedefaultseppunct}\relax
\EndOfBibitem
\bibitem[Casida and Wesolowski(2004)Casida, and Wesolowski]{casi2004}
Casida,~M.~E.; Wesolowski,~T.~A. \emph{{Int. J. Quantum Chem.}} \textbf{2004},
  \emph{96}, 577--588\relax
\mciteBstWouldAddEndPuncttrue
\mciteSetBstMidEndSepPunct{\mcitedefaultmidpunct}
{\mcitedefaultendpunct}{\mcitedefaultseppunct}\relax
\EndOfBibitem
\bibitem[Pavanello(2013)]{pava2013b}
Pavanello,~M. \emph{{J. Chem. Phys.}} \textbf{2013}, \emph{138}, 204118\relax
\mciteBstWouldAddEndPuncttrue
\mciteSetBstMidEndSepPunct{\mcitedefaultmidpunct}
{\mcitedefaultendpunct}{\mcitedefaultseppunct}\relax
\EndOfBibitem
\bibitem[Garc{\'i}a-Lastra \latin{et~al.}(2006)Garc{\'i}a-Lastra, Wesolowski,
  Barriuso, Aramburu, and Moreno]{garc2006}
Garc{\'i}a-Lastra,~J.~M.; Wesolowski,~T.~A.; Barriuso,~M.~T.; Aramburu,~J.~A.;
  Moreno,~M. \emph{{J. Phys.: Condens. Matter}} \textbf{2006}, \emph{18},
  1519--1534\relax
\mciteBstWouldAddEndPuncttrue
\mciteSetBstMidEndSepPunct{\mcitedefaultmidpunct}
{\mcitedefaultendpunct}{\mcitedefaultseppunct}\relax
\EndOfBibitem
\bibitem[Ramos and Pavanello(2016)Ramos, and Pavanello]{ramo2015c}
Ramos,~P.; Pavanello,~M. \emph{{Phys. Chem. Chem. Phys.}} \textbf{2016},
  \emph{18}, 21172\relax
\mciteBstWouldAddEndPuncttrue
\mciteSetBstMidEndSepPunct{\mcitedefaultmidpunct}
{\mcitedefaultendpunct}{\mcitedefaultseppunct}\relax
\EndOfBibitem
\bibitem[Umerbekova \latin{et~al.}(2018)Umerbekova, Zhang, P., and
  Pavanello]{Umerbekova_2018}
Umerbekova,~A.; Zhang,~S.-F.; P.,~S.~K.; Pavanello,~M. \emph{Eur. Phys. J. B}
  \textbf{2018}, \emph{91}\relax
\mciteBstWouldAddEndPuncttrue
\mciteSetBstMidEndSepPunct{\mcitedefaultmidpunct}
{\mcitedefaultendpunct}{\mcitedefaultseppunct}\relax
\EndOfBibitem
\bibitem[Pavanello \latin{et~al.}(2013)Pavanello, {Van Voorhis}, Visscher, and
  Neugebauer]{pava2013a}
Pavanello,~M.; {Van Voorhis},~T.; Visscher,~L.; Neugebauer,~J. \emph{{J. Chem.
  Phys.}} \textbf{2013}, \emph{138}, 054101\relax
\mciteBstWouldAddEndPuncttrue
\mciteSetBstMidEndSepPunct{\mcitedefaultmidpunct}
{\mcitedefaultendpunct}{\mcitedefaultseppunct}\relax
\EndOfBibitem
\bibitem[Solovyeva \latin{et~al.}(2014)Solovyeva, Pavanello, and
  Neugebauer]{solo2014}
Solovyeva,~A.; Pavanello,~M.; Neugebauer,~J. \emph{{J. Chem. Phys.}}
  \textbf{2014}, \emph{140}, 164103\relax
\mciteBstWouldAddEndPuncttrue
\mciteSetBstMidEndSepPunct{\mcitedefaultmidpunct}
{\mcitedefaultendpunct}{\mcitedefaultseppunct}\relax
\EndOfBibitem
\bibitem[T{\"o}lle \latin{et~al.}(2019)T{\"o}lle, Gomes, Ramos, and
  Pavanello]{Toelle_2019}
T{\"o}lle,~J.; Gomes,~A. S.~P.; Ramos,~P.; Pavanello,~M. \emph{Int. J. Quantum
  Chem.} \textbf{2019}, \emph{119}, e25801\relax
\mciteBstWouldAddEndPuncttrue
\mciteSetBstMidEndSepPunct{\mcitedefaultmidpunct}
{\mcitedefaultendpunct}{\mcitedefaultseppunct}\relax
\EndOfBibitem
\bibitem[G{\"o}tz \latin{et~al.}(2009)G{\"o}tz, Beyhan, and Visscher]{gotz2009}
G{\"o}tz,~A.; Beyhan,~S.; Visscher,~L. \emph{{J. Chem. Theory Comput.}}
  \textbf{2009}, \emph{5}, 3161--3174\relax
\mciteBstWouldAddEndPuncttrue
\mciteSetBstMidEndSepPunct{\mcitedefaultmidpunct}
{\mcitedefaultendpunct}{\mcitedefaultseppunct}\relax
\EndOfBibitem
\bibitem[Wesolowski \latin{et~al.}(1996)Wesolowski, Chermette, and
  Weber]{weso1996}
Wesolowski,~T.~A.; Chermette,~H.; Weber,~J. \emph{{J. Chem. Phys.}}
  \textbf{1996}, \emph{105}, 9182\relax
\mciteBstWouldAddEndPuncttrue
\mciteSetBstMidEndSepPunct{\mcitedefaultmidpunct}
{\mcitedefaultendpunct}{\mcitedefaultseppunct}\relax
\EndOfBibitem
\bibitem[Laricchia \latin{et~al.}(2013)Laricchia, Constantin, Fabiano, and
  Sala]{Laricchia_2013b}
Laricchia,~S.; Constantin,~L.~A.; Fabiano,~E.; Sala,~F.~D. \emph{JCTC}
  \textbf{2013}, \emph{10}, 164--179\relax
\mciteBstWouldAddEndPuncttrue
\mciteSetBstMidEndSepPunct{\mcitedefaultmidpunct}
{\mcitedefaultendpunct}{\mcitedefaultseppunct}\relax
\EndOfBibitem
\bibitem[Sinha and Pavanello(2015)Sinha, and Pavanello]{sinh2015}
Sinha,~D.; Pavanello,~M. \emph{{J. Chem. Phys.}} \textbf{2015}, \emph{143},
  084120\relax
\mciteBstWouldAddEndPuncttrue
\mciteSetBstMidEndSepPunct{\mcitedefaultmidpunct}
{\mcitedefaultendpunct}{\mcitedefaultseppunct}\relax
\EndOfBibitem
\bibitem[Schl{\"u}ns \latin{et~al.}(2015)Schl{\"u}ns, Klahr,
  M{\"u}ck-Lichtenfeld, Visscher, and Neugebauer]{schl2015}
Schl{\"u}ns,~D.; Klahr,~K.; M{\"u}ck-Lichtenfeld,~C.; Visscher,~L.;
  Neugebauer,~J. \emph{{Phys. Chem. Chem. Phys.}} \textbf{2015}, \emph{17},
  14323--14341\relax
\mciteBstWouldAddEndPuncttrue
\mciteSetBstMidEndSepPunct{\mcitedefaultmidpunct}
{\mcitedefaultendpunct}{\mcitedefaultseppunct}\relax
\EndOfBibitem
\bibitem[Chac{\'o}n \latin{et~al.}(1985)Chac{\'o}n, Alvarellos, and
  Tarazona]{chac1985}
Chac{\'o}n,~E.; Alvarellos,~J.~E.; Tarazona,~P. \emph{{Phys. Rev. B}}
  \textbf{1985}, \emph{32}, 7868--7877\relax
\mciteBstWouldAddEndPuncttrue
\mciteSetBstMidEndSepPunct{\mcitedefaultmidpunct}
{\mcitedefaultendpunct}{\mcitedefaultseppunct}\relax
\EndOfBibitem
\bibitem[Wesolowski and Wang(2013)Wesolowski, and Wang]{wesolowski2013recent}
Wesolowski,~T.~A.; Wang,~Y.~A. \emph{Recent progress in orbital-free density
  functional theory}; World Scientific, 2013; Vol.~6\relax
\mciteBstWouldAddEndPuncttrue
\mciteSetBstMidEndSepPunct{\mcitedefaultmidpunct}
{\mcitedefaultendpunct}{\mcitedefaultseppunct}\relax
\EndOfBibitem
\bibitem[Karasiev and Trickey(2012)Karasiev, and Trickey]{karasiev2012issues}
Karasiev,~V.~V.; Trickey,~S.~B. \emph{Comput. Phys. Commun.} \textbf{2012},
  \emph{183}, 2519--2527\relax
\mciteBstWouldAddEndPuncttrue
\mciteSetBstMidEndSepPunct{\mcitedefaultmidpunct}
{\mcitedefaultendpunct}{\mcitedefaultseppunct}\relax
\EndOfBibitem
\bibitem[Witt \latin{et~al.}(2018)Witt, Beatriz, Dieterich, and
  Carter]{witt2018orbital}
Witt,~W.~C.; Beatriz,~G.; Dieterich,~J.~M.; Carter,~E.~A. \emph{J. M. Res.}
  \textbf{2018}, \emph{33}, 777--795\relax
\mciteBstWouldAddEndPuncttrue
\mciteSetBstMidEndSepPunct{\mcitedefaultmidpunct}
{\mcitedefaultendpunct}{\mcitedefaultseppunct}\relax
\EndOfBibitem
\bibitem[Nafziger and Wasserman(2015)Nafziger, and Wasserman]{Nafziger_2015}
Nafziger,~J.; Wasserman,~A. \emph{{J. Chem. Phys.}} \textbf{2015}, \emph{143},
  234105\relax
\mciteBstWouldAddEndPuncttrue
\mciteSetBstMidEndSepPunct{\mcitedefaultmidpunct}
{\mcitedefaultendpunct}{\mcitedefaultseppunct}\relax
\EndOfBibitem
\bibitem[Pavanello and Neugebauer(2011)Pavanello, and Neugebauer]{pava2011b}
Pavanello,~M.; Neugebauer,~J. \emph{{J. Chem. Phys.}} \textbf{2011},
  \emph{135}, 234103\relax
\mciteBstWouldAddEndPuncttrue
\mciteSetBstMidEndSepPunct{\mcitedefaultmidpunct}
{\mcitedefaultendpunct}{\mcitedefaultseppunct}\relax
\EndOfBibitem
\bibitem[Ramos \latin{et~al.}(2016)Ramos, Mankarious, and Pavanello]{ramo2015b}
Ramos,~P.; Mankarious,~M.; Pavanello,~M. \emph{{Practical Aspects in
  Computational Chemistry IV}}; Springer, 2016; Chapter 4\relax
\mciteBstWouldAddEndPuncttrue
\mciteSetBstMidEndSepPunct{\mcitedefaultmidpunct}
{\mcitedefaultendpunct}{\mcitedefaultseppunct}\relax
\EndOfBibitem
\bibitem[Fermi(1927)]{fermi1927}
Fermi,~E. \emph{Rend. Accad. Naz. Lincei} \textbf{1927}, \emph{6},
  602--607\relax
\mciteBstWouldAddEndPuncttrue
\mciteSetBstMidEndSepPunct{\mcitedefaultmidpunct}
{\mcitedefaultendpunct}{\mcitedefaultseppunct}\relax
\EndOfBibitem
\bibitem[Thomas(1927)]{thom1927}
Thomas,~L.~A. \emph{Proc. Camb. Phil. Soc.} \textbf{1927}, \emph{23},
  542--548\relax
\mciteBstWouldAddEndPuncttrue
\mciteSetBstMidEndSepPunct{\mcitedefaultmidpunct}
{\mcitedefaultendpunct}{\mcitedefaultseppunct}\relax
\EndOfBibitem
\bibitem[{von Weizs{\"a}cker}(1935)]{weiz1935}
{von Weizs{\"a}cker},~C.~F. \emph{Z. Physik} \textbf{1935}, \emph{96},
  431--458\relax
\mciteBstWouldAddEndPuncttrue
\mciteSetBstMidEndSepPunct{\mcitedefaultmidpunct}
{\mcitedefaultendpunct}{\mcitedefaultseppunct}\relax
\EndOfBibitem
\bibitem[Wang and Carter(2000)Wang, and Carter]{wang2000}
Wang,~Y.~A.; Carter,~E.~A. In \emph{{Theoretical Methods in Condensed Phase
  Chemistry}}; Schwartz,~S.~D., Ed.; Kluwer: Dordrecht, 2000; pp 117--184\relax
\mciteBstWouldAddEndPuncttrue
\mciteSetBstMidEndSepPunct{\mcitedefaultmidpunct}
{\mcitedefaultendpunct}{\mcitedefaultseppunct}\relax
\EndOfBibitem
\bibitem[Wang \latin{et~al.}(1998)Wang, Govind, and Carter]{wang1998}
Wang,~Y.~A.; Govind,~N.; Carter,~E.~A. \emph{{Phys. Rev. B}} \textbf{1998},
  \emph{58}, 13465--13471\relax
\mciteBstWouldAddEndPuncttrue
\mciteSetBstMidEndSepPunct{\mcitedefaultmidpunct}
{\mcitedefaultendpunct}{\mcitedefaultseppunct}\relax
\EndOfBibitem
\bibitem[Wang \latin{et~al.}(1999)Wang, Govind, and Carter]{wang1999}
Wang,~Y.~A.; Govind,~N.; Carter,~E.~A. \emph{{Phys. Rev. B}} \textbf{1999},
  \emph{60}, 16350--16358\relax
\mciteBstWouldAddEndPuncttrue
\mciteSetBstMidEndSepPunct{\mcitedefaultmidpunct}
{\mcitedefaultendpunct}{\mcitedefaultseppunct}\relax
\EndOfBibitem
\bibitem[Huang and Carter(2010)Huang, and Carter]{huan2010}
Huang,~C.; Carter,~E.~A. \emph{{Phys. Rev. B}} \textbf{2010}, \emph{81},
  045206\relax
\mciteBstWouldAddEndPuncttrue
\mciteSetBstMidEndSepPunct{\mcitedefaultmidpunct}
{\mcitedefaultendpunct}{\mcitedefaultseppunct}\relax
\EndOfBibitem
\bibitem[Wang and Teter(1992)Wang, and Teter]{wang1992}
Wang,~L.-W.; Teter,~M.~P. \emph{{Phys. Rev. B}} \textbf{1992}, \emph{45},
  13196--13220\relax
\mciteBstWouldAddEndPuncttrue
\mciteSetBstMidEndSepPunct{\mcitedefaultmidpunct}
{\mcitedefaultendpunct}{\mcitedefaultseppunct}\relax
\EndOfBibitem
\bibitem[Mi \latin{et~al.}(2018)Mi, Genova, and Pavanello]{mi2018nonlocal}
Mi,~W.; Genova,~A.; Pavanello,~M. \emph{{J. Chem. Phys.}} \textbf{2018},
  \emph{148}, 184107\relax
\mciteBstWouldAddEndPuncttrue
\mciteSetBstMidEndSepPunct{\mcitedefaultmidpunct}
{\mcitedefaultendpunct}{\mcitedefaultseppunct}\relax
\EndOfBibitem
\bibitem[Pearson \latin{et~al.}(1993)Pearson, Smargiassi, and
  Madden]{Pearson_1993}
Pearson,~M.; Smargiassi,~E.; Madden,~P.~A. \emph{{J. Phys.: Condens. Matter}}
  \textbf{1993}, \emph{5}, 3221--3240\relax
\mciteBstWouldAddEndPuncttrue
\mciteSetBstMidEndSepPunct{\mcitedefaultmidpunct}
{\mcitedefaultendpunct}{\mcitedefaultseppunct}\relax
\EndOfBibitem
\bibitem[Smargiassi and Madden(1994)Smargiassi, and Madden]{smar1994}
Smargiassi,~E.; Madden,~P.~A. \emph{{Phys. Rev. B}} \textbf{1994}, \emph{49},
  5220--5226\relax
\mciteBstWouldAddEndPuncttrue
\mciteSetBstMidEndSepPunct{\mcitedefaultmidpunct}
{\mcitedefaultendpunct}{\mcitedefaultseppunct}\relax
\EndOfBibitem
\bibitem[Perrot(1994)]{perr1994}
Perrot,~F. \emph{{J. Phys.: Condens. Matter}} \textbf{1994}, \emph{6},
  431--446\relax
\mciteBstWouldAddEndPuncttrue
\mciteSetBstMidEndSepPunct{\mcitedefaultmidpunct}
{\mcitedefaultendpunct}{\mcitedefaultseppunct}\relax
\EndOfBibitem
\bibitem[Mi and Pavanello(2019)Mi, and Pavanello]{mi2019LMGP}
Mi,~W.; Pavanello,~M. \emph{Phys. Rev. B Rapid Commun.} \textbf{2019},
  \emph{100}, 041105\relax
\mciteBstWouldAddEndPuncttrue
\mciteSetBstMidEndSepPunct{\mcitedefaultmidpunct}
{\mcitedefaultendpunct}{\mcitedefaultseppunct}\relax
\EndOfBibitem
\bibitem[Perdew \latin{et~al.}(1996)Perdew, Burke, and Ernzerhof]{PBEc}
Perdew,~J.~P.; Burke,~K.; Ernzerhof,~M. \emph{{Phys. Rev. Lett.}}
  \textbf{1996}, \emph{77}, 3865--3868\relax
\mciteBstWouldAddEndPuncttrue
\mciteSetBstMidEndSepPunct{\mcitedefaultmidpunct}
{\mcitedefaultendpunct}{\mcitedefaultseppunct}\relax
\EndOfBibitem
\bibitem[Grafova \latin{et~al.}(2010)Grafova, Pitonak, Rezac, and
  Hobza]{Grafova_2010}
Grafova,~L.; Pitonak,~M.; Rezac,~J.; Hobza,~P. \emph{{J. Chem. Theory Comput.}}
  \textbf{2010}, \emph{6}, 2365--2376\relax
\mciteBstWouldAddEndPuncttrue
\mciteSetBstMidEndSepPunct{\mcitedefaultmidpunct}
{\mcitedefaultendpunct}{\mcitedefaultseppunct}\relax
\EndOfBibitem
\bibitem[Giannozzi \latin{et~al.}(2009)Giannozzi, Baroni, Bonini, Calandra,
  Car, Cavazzoni, Ceresoli, Chiarotti, Cococcioni, Dabo, {Dal Corso},
  de~Gironcoli, Fabris, Fratesi, Gebauer, Gerstmann, Gougoussis, Kokalj,
  Lazzeri, Martin-Samos, Marzari, Mauri, Mazzarello, Paolini, Pasquarello,
  Paulatto, Sbraccia, Scandolo, Sclauzero, Seitsonen, Smogunov, Umari, and
  Wentzcovitch]{qe}
Giannozzi,~P. \latin{et~al.}  \emph{J. Phys.: Cond. Mat.} \textbf{2009},
  \emph{21}, 395502\relax
\mciteBstWouldAddEndPuncttrue
\mciteSetBstMidEndSepPunct{\mcitedefaultmidpunct}
{\mcitedefaultendpunct}{\mcitedefaultseppunct}\relax
\EndOfBibitem
\bibitem[Sabatini \latin{et~al.}(2013)Sabatini, Gorni, and De~Gironcoli]{rvv10}
Sabatini,~R.; Gorni,~T.; De~Gironcoli,~S. \emph{Physical Review B}
  \textbf{2013}, \emph{87}, 041108\relax
\mciteBstWouldAddEndPuncttrue
\mciteSetBstMidEndSepPunct{\mcitedefaultmidpunct}
{\mcitedefaultendpunct}{\mcitedefaultseppunct}\relax
\EndOfBibitem
\bibitem[Vanderbilt(1990)]{vand1990}
Vanderbilt,~D. \emph{Phys. Rev. B} \textbf{1990}, \emph{41}, 7892--7895\relax
\mciteBstWouldAddEndPuncttrue
\mciteSetBstMidEndSepPunct{\mcitedefaultmidpunct}
{\mcitedefaultendpunct}{\mcitedefaultseppunct}\relax
\EndOfBibitem
\bibitem[Garrity \latin{et~al.}(2014)Garrity, Bennett, Rabe, and
  Vanderbilt]{garr2014}
Garrity,~K.~F.; Bennett,~J.~W.; Rabe,~K.~M.; Vanderbilt,~D. \emph{Computational
  Materials Science} \textbf{2014}, \emph{81}, 446--452\relax
\mciteBstWouldAddEndPuncttrue
\mciteSetBstMidEndSepPunct{\mcitedefaultmidpunct}
{\mcitedefaultendpunct}{\mcitedefaultseppunct}\relax
\EndOfBibitem
\bibitem[epa()]{epaps}
{See Supplementary Material Document at [URL will be inserted by publisher] for
  additional tables and figures.}\relax
\mciteBstWouldAddEndPunctfalse
\mciteSetBstMidEndSepPunct{\mcitedefaultmidpunct}
{}{\mcitedefaultseppunct}\relax
\EndOfBibitem
\bibitem[Raghavachari \latin{et~al.}(1989)Raghavachari, Trucks, Pople, and
  Head-Gordon]{CCSDT}
Raghavachari,~K.; Trucks,~G.~W.; Pople,~J.~A.; Head-Gordon,~M. \emph{Chemical
  Physics Letters} \textbf{1989}, \emph{157}, 479--483\relax
\mciteBstWouldAddEndPuncttrue
\mciteSetBstMidEndSepPunct{\mcitedefaultmidpunct}
{\mcitedefaultendpunct}{\mcitedefaultseppunct}\relax
\EndOfBibitem
\bibitem[Vydrov and {Van Voorhis}(2012)Vydrov, and {Van Voorhis}]{vydr2011}
Vydrov,~O.~A.; {Van Voorhis},~T. \emph{{J. Chem. Theory Comput.}}
  \textbf{2012}, \emph{8}, 1929--1934\relax
\mciteBstWouldAddEndPuncttrue
\mciteSetBstMidEndSepPunct{\mcitedefaultmidpunct}
{\mcitedefaultendpunct}{\mcitedefaultseppunct}\relax
\EndOfBibitem
\bibitem[Medvedev \latin{et~al.}(2017)Medvedev, Bushmarinov, Sun, Perdew, and
  Lyssenko]{Medvedev_2017}
Medvedev,~M.~G.; Bushmarinov,~I.~S.; Sun,~J.; Perdew,~J.~P.; Lyssenko,~K.~A.
  \emph{Science} \textbf{2017}, \emph{355}, 49--52\relax
\mciteBstWouldAddEndPuncttrue
\mciteSetBstMidEndSepPunct{\mcitedefaultmidpunct}
{\mcitedefaultendpunct}{\mcitedefaultseppunct}\relax
\EndOfBibitem
\bibitem[Wasserman \latin{et~al.}(2017)Wasserman, Nafziger, Jiang, Kim, Sim,
  and Burke]{Wasserman_2017}
Wasserman,~A.; Nafziger,~J.; Jiang,~K.; Kim,~M.-C.; Sim,~E.; Burke,~K.
  \emph{Annual Review of Physical Chemistry} \textbf{2017}, \emph{68},
  555--581\relax
\mciteBstWouldAddEndPuncttrue
\mciteSetBstMidEndSepPunct{\mcitedefaultmidpunct}
{\mcitedefaultendpunct}{\mcitedefaultseppunct}\relax
\EndOfBibitem
\end{mcitethebibliography}

\end{document}